\documentclass{ws-gm}

\usepackage[english]{babel}
\usepackage[utf8]{inputenc}
\usepackage{tikz}
\usepackage{tikz-cd}

\usepackage{amsmath}
\usepackage{mathtools}
\usepackage{mathrsfs}
\usepackage{empheq}

\usepackage[sort,compress]{cite}

\usepackage{xcolor}
\usepackage[verbose]{hyperref}
\hypersetup{colorlinks=false,allbordercolors=blue,pdfborderstyle={/S/U/W 1}}
\newcommand{\R}{\mathbb{R}}
\newcommand{\Z}{\mathbb{Z}}
\newcommand{\SmoothSet}{\mathsf{SmoothSet}}
\newcommand{\Mfd}{\mathsf{Mfd}}
\newcommand{\Alg}{\mathsf{Alg}}
\newcommand{\Set}{\mathsf{Set}}
\newcommand{\Eucl}{\mathsf{Eucl}}

\begin{document}

\markboth{A. Ibort, A. Mas}{Smooth sets of fields}

\catchline{}{}{}{}{}
%
%%%%%%%%%%%%%%%%%%%%%%%%%%%%%%%%%%%%%%%%%%%%%%%%%%%%%%%%%%%%%%%%%%%%

\title{Smooth sets of fields: A pedagogical introduction}

\author{Alberto Ibort}

\address{Department of Mathematics, Universidad Carlos III de Madrid, Avda. de la Universidad 30\\
Leganés, 28911 Madrid, and ICMAT, Calle Nicolas Cabrera, 13-15. Campus de Cantoblanco, 28049 Madrid, Spain \\
\email{albertoi@math.uc3m.es}}

\author{Arnau Mas}

\address{Department of Mathematics, Universidad Carlos III de Madrid, Avda. de la Universidad 30\\
Legan\'es, 28911 Madrid, and ICMAT, Calle Nicolas Cabrera, 13-15. Campus de Cantoblanco, 28049 Madrid, Spain\\
\email{amas@math.uc3m.es}}

\maketitle

\begin{history}
\received{(Day Month Year)}
\revised{(Day Month Year)}
\accepted{(Day Month Year)}
\published{(Day Month Year)}
\comby{(Handling Editor)} % Communicated by optional
\end{history}

\begin{abstract}
In order to provide a good categorical setting to the many different spaces of fields arising in the description of physical theories, a pedagogical introduction to the categorical notion of smooth sets following the steps of Giotopoulos and Sati (J. Geom. Phys. {\bf 213}, 2025) is provided and some simple properties of the topos of smooth sets are discussed. The introduction of geometrical structures into such spaces is illustrated via the specific examples of the tangent functor and the variational bicomplex.
 \end{abstract}



%%%%%%%%%%%
%%%%%%%%%%%

\section{Introduction}

Classical fields $\phi$ are described mathematically as sections of a given bundle $\pi \colon E \to M$, where $M$ is a manifold\footnote{Here and in all that follows ``manifold'' will refer to a smooth second countable manifold.} that often models the spacetime of the theory (as in the case of relativistic field theories), or a manifold carrying additional geometric structures (like Riemann surfaces in the case of string theories, or higher dimensional Riemannian manifolds in Euclidean field theories for instance).     The bundle $E$ is typically a vector bundle, a principal bundle or a bundle with standard fibre a manifold carrying some geometrical structure  (like a Poisson structure in the case of Poisson sigma models).     Thus the bundle $\pi \colon E \to M$ provides the auxiliary geometrical background upon which we construct the theory we are interested in. 

The geometrical background provided by the bundle $E$ constitutes the departing point for the construction of a dynamical theory of fields that, in its Lagrangian picture, uses the corresponding jet bundles $J^k\pi \colon J^kE \to E$, $k \geq 0$, \cite{Saunders1989} typically the first jet bundle for first order field theories, while the multisymplectic structures carried on by the affine duals of such bundles, lead to the Hamiltonian covariant picture of field theories (see, for instance, \cite{Carinena1991, Ibort2017}, and references therein).    In spite of these neat geometrical descriptions, the truth is that the space of fields themselves, either the sets of smooth sections of the relevant bundles or the sets of solutions of the dynamical equations of the theory (generically called in what follows Euler-Lagrange equations) fail to be smooth manifolds (finite or infinite dimensional).    Indeed, the space of smooth functions on a finite-dimensional manifold is just a Fr\'echet space (a complete, metrizable and locally convex topological space), when equipped with the natural topologies inherited from the background manifolds.    

In spite of this, the geometry carried by the underlying spaces percolates to the spaces of functions defined on them in a natural way.   For instance, if $\pi \colon E \to M$ is a Hermitian bundle over the Riemannian manifold, then the space of sections $\Gamma (E)$ carries a natural inner product $\langle \phi, \psi \rangle = \int_M \langle \phi (x), \psi (x) \rangle_x d\mu (x)$, where $\langle \cdot, \cdot \rangle_x$ denotes the inner product along the fibre $E_x = \pi^{-1}(x)$ of the bundle $E$, and $\mu$ is the measure defined by the Riemannian metric on $M$.    Even more, if we consider variations of a given section $\pi$ as defined by smooth maps $\Phi \colon (-\epsilon, \epsilon) \times M \to E$, $\Phi (s,x) = \phi_s(x)$, $\phi_s$ a cross section for each $s$, then, the vertical vector field $\delta\phi$ on $E$ along $\phi$ defined by $\delta\phi (x) = \frac{\partial}{\partial s} \Phi (s,x)\mid_{s = 0}$, can be thought as a tangent vector  at $\phi$ to the space $\Gamma (E)$ of fields, and the Hermitian structure on the underlying bundle defines a natural metric $\langle \delta_1\phi, \delta_2\phi \rangle_\phi = \int_M \langle \delta_1\phi(x), \delta_2\phi (x) \rangle_x d\mu(x)$ on it.     Thus, geometrical structures on the underlying spaces used to construct the theory get promoted to ``geometrical'' structures on the corresponding spaces of fields.     

A most relevant example of this phenomena is given by the so called Peierls bracket in the space of solutions of Euler-Lagrange equations.   Indeed, in this case, the space of solutions of Euler-Lagrange equations, again an infinite-dimensional Fréchet space which is not a linear space whenever the Euler-Lagrange equations are non-linear, carries a canonical presymplectic structure, i.e., a closed but in general degenerate, 2-form $\Omega$ (see \cite{Ciaglia2024} and refs. therein).   Of course because the space of solutions of Euler-Lagrange equations is not in general a smooth manifold, the previous statement must be interpreted in a similar way as we were doing before for the space of sections of a vector bundle and declaring what are the ``tangent vectors'' to such space.   

Although in many physical applications an \textit{ad hoc} description of the geometry of the spaces of fields like the ones sketched before is enough for most practical purposes, there is always a moment when it is necessary to understand in what sense the arguments based on such constructions are precise and well stated beyond a tenuous principle of analogy between finite-dimensional differential geometry and the induced geometries on spaces of fields.      

The previous observation is of utmost importance when, for instance, one is trying to uplift arguments from symmetry theory to the infinite-dimensional setting.  The well developed machinery of finite dimensional symplectic geometry can be extended to handle symmetries in infinite dimensional situations.   This sort of arguments have been forcefully stressed by the observation that, for instance, Duistermaat-Heckman convexity integration theory can be extended to the case of 2-dimensional gauge theories (see, for instance, \cite{Witten1992}) or by realizing the role of Atiyah's index theorem as the geometry of manifolds of loops \cite{Atiyah1985}.      The need for a better understanding of the geometry of infinite dimensional spaces of fields becomes more relevant when we try to use them to address problems in low dimensional topology or symplectic geometry by using topological field theories.

A natural way to tackle these situations that has been exploited since long ago consists in completing the spaces of fields providing the appropriate analytical skeleton that makes them smooth manifolds modelled on Hilbert of Banach spaces.  The spaces of fields of interest, for instance spaces of connections in gauge theories, are completed using different Sobolev norms so that they become Hilbert manifolds (see, for instance,  \cite{Asorey1981,Mitter1980}) or, in a different class of applications, the spaces of diffeomorphisms of a given manifold are completed into what are called inductive Lie Banach/Hilbert groups \cite{Omori1974}, leading to important results in the theory of solutions of Euler's equations for an incompressible fluid or other results in general relativity \cite{Ebin1970, Anderson1997, Chernoff1974}.  This was also the path followed, for instance, in \cite{Ciaglia2024} to provide a sound analytical background to the study of the geometrical structure of the Peierls bracket.  

The heavy use of analysis involved in the previous approaches was considered unnecessary by other schools of geometers that were convinced of the prevalence of the geometrical approach over the fuzz introduced by the use of Hilbert spaces and other topological completions.     This point of view led to the construction of a purely geometric theory of partial differential equations started in the works of Spencer, Goldschmidt, Garcia, etc., (see, for instance, \cite{Spencer1962, Goldschmidt1967, Garcia1977}) and encoded today in the so called geometry of the variational bicomplex associated to a given bundle.    In other words, given a bundle $\pi \colon E \to M$, the collection of all jet bundles $J^k \pi$, $k \in \mathbb{N}$, defines a geometrical object, denoted by $J^\infty \pi$, carrying a canonical Cartan distribution.  Then, it is possible to construct a double complex of forms on $J^\infty\pi$ called the variational bicomplex whose structure plays a fundamental role in the study of the properties of field theories whose Euler-Lagrange equations are partial differential equations.  Partial differential equations determine subspaces of $J^\infty \pi$ and their properties can be studied by means of the natural structures constructed on $J^\infty\pi$.    Infinite dimensional spaces constructed as the infinite prolongation of partial differential equations carrying a Cartan distribution were called ``diffieties'' by A. Vinogradov, and it was shown how the cohomology of the spectral sequences associated to the double complex of forms on it could be used to analyse the structure of the given equations (see, for instance, \cite{Vinogradov2001} and references therein).   

In a parallel road, J.M. Souriau \cite{Souriau1980} introduced a completely different idea to handle spaces carrying geometrical structures.  Souriau's main insight consisted on relaxing the notion of local charts used in the definition of smooth manifolds but preserving the right properties allowing local constructions and glueing. This extension of the notion of local charts are called ``plots''.    Plots do not have to provide a local identification of a manifold with an open set of $\mathbb{R}^n$, but they only have to behave nicely in a categorical way, essentially they have nice local properties that allows us to localize and glue objects and they must transform properly under change of local parameters. Such spaces are called diffeological spaces. 
\begin{definition}\label{def:diffeological}
A diffeological space is a set $X$ equipped with a family of maps $f \colon U \to X$, called plots, where $U$ are open subsets in $\mathbb{R}^n$, $n \in \mathbb{N}$, satisfying: 
\begin{itemize}
\item[a)] Locality.  If $f \colon U \cup V \to X$ is a map such that $f\mid_U$ and $f\mid_V$ are plots, then $f$ is a plot.
\item[b)] Smooth compatibility.  If   $f \colon U \to X$ is a plot and $\varphi \colon V \to U$ is a smooth map, with $V$ and open set in $\mathbb{R}^m$, then $f \circ \varphi \colon V \to X$ is also a plot and, 
\item[c)] every constant map is a plot. 
\end{itemize}
\end{definition}
 The previous conditions can be expressed in a condensed way by saying that a diffeological space is a concrete sheaf on the site of open subsets on $\mathbb{R}^n$ ($n \in \mathbb{N}$) (see later on, Sect. \ref{sec:smooth}, for a detailed discussion of these notions).   

Diffeological spaces have proved to be very helpful to understand a number of issues that are badly dealt with in the category of manifolds.     Certainly, apart from the intrinsic analytical difficulties associated to infinite dimensional spaces of maps mentioned above, the differential geometry of smooth manifolds is intrinsically unable to deal with other problems that arise often in the work of the practitioner and that are related to the emergence of singularities associated to quotients, foliations and other constructions with a natural geometric flavor. Besides diffeological spaces, a number of variations on the notion of smooth manifold were introduced to cope with these difficulties. As a non-exhaustive list, $C^\infty$-differentiable spaces  (in \cite{NavarroGonzalez2003}), Fr\"olicher spaces or Chen spaces (see \cite{Michor1997,Baez2011}). Diffeological spaces provide an adequate setting to deal with these situations (see \cite{IglesiasZemmour2013} and references therein for an exhaustive discussion of these aspects).     However, as it will discussed at length afterwards, a natural extension of the notion of diffeological spaces that allows for more general choices of plots to ``probe'' the system, called smooth sets \cite{Schreiber2023}, will be our choice to deal with crucial physical applications like fermion fields, infinitesimal structures, and higher gauge fields (see, for instance  \cite{Giotopoulos2025b}).

It was acknowledged since a long time that the category of smooth manifolds $\mathsf{Mfd}$, that is, the category whose objects are smooth manifolds and whose morphisms are smooth maps between manifolds, is ``too small'' and not well behaved (see, for instance, \cite{Ehresmann1963,Pradines2007}) in a categorical way.   Basic 
constructions  in category theory, like pull-backs, quotients and exponentials, fail to be well defined in the category $\mathsf{Mfd}$.   In general the quotient of a smooth manifold by an equivalence relation is not a smooth manifold anymore and subsets of smooth manifolds are not manifolds.   In addition to it, the set of morphisms from $M$ to $N$ fails to be a manifold.     Many of these operations are defined as limits and colimits in categorical terms and, as it turns out, a Cartesian closed category where all limits and colimits exist is called a topos (see Sects. \ref{sec:topos}, \ref{sec:geometry}, later on for detailed explanations).  For instance the category $\mathsf{Set}$ whose objects are sets and morphisms are maps among sets, is a topos, however the category $\mathsf{Mfd}$ is not.    

Of course even if we are working in a topos category, it is not a minor task to extend the basic notions of finite-dimensional geometry to it or, in other words, to provide the framework to do differential geometry in such category.    The systematic way do it carries the name ``synthetic differential geometry'' (see \cite{Kock2010} and references therein) but this will not be the approach we will take here.     As it happens there is a natural topos containing all the notions of geometrical spaces mentioned before.  Such category is called the topos of ``smooth spaces'' and contains the category of diffeological spaces $\mathsf{Diff}$ as a full subcategory and it will be denoted by $\mathsf{SmoothSet}$ (see, for instance \cite{Baez2011,Schreiber2023} and references therein).    We are willing to join the efforts to provide a more natural description of the geometrical structures underlying spaces of fields by means of the categorical notions brought up by the theory of topos and more specifically the theory of smooth spaces.

Thus the present paper aims to offer a pedagogical introduction to the notion of smooth spaces following the steps of \cite{Schreiber2023,Giotopoulos2025}, and provide a few insights on its usefulness describing the geometry of spaces of fields.  We will show in what sense various spaces of fields arising in the description of field theories are smooth spaces.    To do that we will follow a double path, on one side explicit smooth spaces structures for given spaces of fields will be considered (mostly diffeological spaces) and, in a more abstract and powerful way, it will be shown that certain spaces are smooth sets by using abstract properties of the topos of smooth sets.  For instance, the infinite jet bundle $J^\infty\pi$ is a smooth space and many structures associated to it can be discussed in this way.   In doing so we will pave the way to construct a categorical description of the variational bicomplex and the abstract description of field theories associated to it.    

The article will be organized as follows.    The basic notions leading to the notion of smooth spaces as well as the most important properties of such category will be discussed in Sect. \ref{sec:smooth_gen}.  Section \ref{sec:examples} will be devoted to discuss in detail a few significant examples: spaces of sections of bundles, gauge field theories, the groupoid of evolutions in general relativity and superspaces. In Sect. \ref{sec:geometry} some properties of the topos structure of the category $\mathsf{SmoothSet}$ will be used to address geometrical problems and, in Sect. \ref{sec:bicomplex}, the categorical background of the variational bicomplex will be analyzed and a few simple consequences will be highlighted. 

%Finally, some hints on a higher geometry of field theories will be discussed in Sect. \ref{sec:higher}

%%%%%%%%%%%
%%%%%%%%%%%

\section{Smooth sets}\label{sec:smooth_gen}
\subsection{Smooth sets: basic notions}\label{sec:smooth}

As it was discussed in the introduction the category of smooth manifolds is not a particularly well-behaved category. The following are a number of constructions that fail to be manifolds except in a number of particular situations:
  \begin{itemize}
		\item The quotient $M/G$ of a manifold \( M \) by the action of a Lie group \( G \) will not be a manifold unless the action of \( G \) is free and proper. As examples of the pathologies that can occur, consider the action of \( SO(2) \) on the plane by rotations around the origin. This is not free because the origin is fixed, and the image of the origin in the quotient has no neighbourhood homeomorphic to an open subset of \( \R^2 \). Another paradigmatic example is the action of \( \Z \) on \( S^1 \) by an irrational rotation. This is free, but not proper. Every orbit is dense in \( S^1 \) so that the quotient has the chaotic topology. 
		\item Given two smooth maps \( f \colon M \to Z \) and \( g \colon N \to Z \), the pullback \( M \times_Z N = \{ (x,y) \in M \times N \mid f(x) = g(y) \} \) will not be a manifold in general (a sufficient condition is that \( f \) and \( g \) are transverse).
		\item the space of smooth maps \( C^\infty(M,N) \) between two manifolds will not be a manifold (if the dimensions of \( M \) and \( N \) are both non-zero), or at least certainly not a finite dimensional one. They are Fréchet manifolds\footnote{A Fréchet manifold is a topological space locally homeomeomorphic to a Fréchet space.}, but this is not particularly useful since we are rarely interested in analytical questions of convergence and uniformity when dealing with field theory.
  \end{itemize}

It is our goal here to describe how the tools of category theory (the language of sheaves and topoi in particular) naturally lead to a solution to this problem, by providing a category into which the category \( \Mfd \) of smooth manifolds embeds fully faithfully\footnote{We could embed \( \Mfd \) inside the catgory of sets, which is a topos, by forgetting the differentiable structure, but this fails to be a full embedding, because there are many more maps between the sets underlying two manifolds than there are smooth maps between them.}, and which also contains all of the constructions outlined above (i.e., it is a topos). But more important, the way we motivate this construction is very fundamental and deeply rooted in insights from physics. From just two fundamental axioms, we can bootstrap the definition of the \emph{topos of smooth sets}. 

The guiding assumption, which has been known to physicists for a long time, is that to know properties of some object we probe it with test particles. The objects to be probed will turn out to be the objects in the category we are trying to pin down. The ``test particles'' or ``probes'' will be objects in some simple category of domains. Asking that this probing happens in a way that is consistent with the structure of the category of probes will inevitable lead to the following conclusion:
\begin{definition}\label{def:smooth}
A smooth set $X$ is  a sheaf on the site of open subsets of Euclidean spaces.
\end{definition}

We spend the rest of this section unwinding the previous definition. Let \( X \) be the ``space'' we wish to probe.  It is relevant to stress here that $X$ does not have to be a set.  The smooth set $X$ is determined by assigning sets to open sets in Euclidean spaces, that is, for each open set \( U \subseteq \R^n \), for any \( n \), all we assume is that there exists a set of all the ways in which we can ``probe'' \( X \) with \( U \), which we write \( X(U) \). We call \( X(U) \) the set of \( U \)-shaped plots of \( X \) and its elements $f \in X(U)$ plots.  If \( U \) were a single point then \( X(U) \) would be interpreted as the set of all the ways in which we can single out a point of \( X \), i.e. the set of points of \( X \)\footnote{Of course, stated in this way this does not make sense as a point is not an open set in $\mathbb{R}^n$, however it is possible to do that making full use of the topos structure and the so called universal subobject classifier \cite{Leinster2011}.  Note again that it is not always the case that a smooth set is its underlying set of points equipped with additional structure, when this happens we say the smooth set is concrete, see \ref{ex:non concrete}}.  If \( U \) is an interval, then \( X(U) \) would be interpreted as the set of all the ways in which we can lay an interval inside of \( U \), etc. 

A smooth set \( X \) then defines some assignment from the collection of all open subsets of every \( \R^n \), which are the objects of the Euclidean category \( \Eucl \) whose morphisms are just smooth maps $\phi \colon U \to V$, to the collection of all sets \( \Set \). Now, given some smooth map \( \phi \colon U \to V \), we can use it to transform \( V \)-shaped plots of \( X \) into \( U \)-shaped plots of \( X \) --this is not by precomposition with \( \phi \), since plots $f \in X(U)$ are not necessarily functions defined on $U$, but it is a good picture to have in mind. Hence, \( \phi \) is lifted through \( X \) into a map of sets of plots
\begin{equation}
	X(\phi) \colon X(V) \to X(U),
\end{equation}
which crucially reverses directions. But if we have two such smooth maps \( \phi \colon U \to V \) and \( \psi \colon V \to W \), this operation can happen in two different ways. We could first compose \( \psi \) after \( \phi \) and then apply \( X \) to the result, or first apply \( X \) and then compose the result. If we want the probing of \( X \) to happen consistently, these two operations had better coincide. Thus we require the following equality
  \begin{equation}\label{eq:consistency}
    X(\psi \circ \phi) = X(\phi) \circ X(\psi).
  \end{equation}
This is exactly the statement that the assignment from domains to plots defined by \( X \) is in fact a contravariant functor from the category \( \Eucl \) of open domains and \( \Set \) the category of sets. The assignment $X$ is a functor because it respects composition, but contravariant because it reverses directions, Eq. (\ref{eq:consistency}). A contravariant functor whose codomain is \( \Set \) is also called a \emph{presheaf}.   A contravariant functor $F$ on the category $\mathsf{C}$ with codomain the category $\mathsf{D}$, is a functor $F \colon \mathsf{C}^\mathrm{op} \to \mathsf{D}$, where $\mathsf{C}^\mathrm{op}$, denotes the opposite category to $\mathsf{C}$, that is, the category with the same objects but all morphisms reversed.  Thus the presheaf $X$ is a functor $X \colon \mathsf{Eucl}^\mathrm{op} \to \mathsf{Set}$.

\begin{figure}
\centerline{\includegraphics[width=12cm]{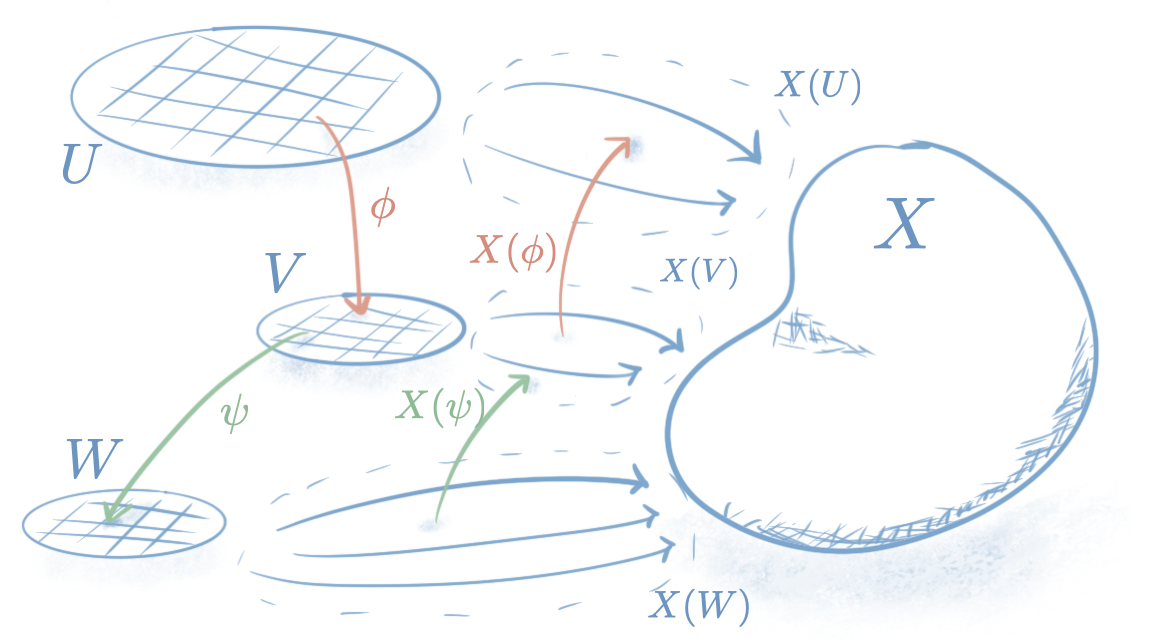}}
\vspace*{8pt}
\caption{A schematic illustration of a smooth set.  The sets $X(U)$ denote the sets of plots of shape $U$, and the maps $X(\phi)$ determine how these plots change when we change the domain of reference $\varphi \colon U \to V$.}\label{fig:smoothset}
\end{figure}

\begin{figure}
\centerline{\includegraphics[width=12cm]{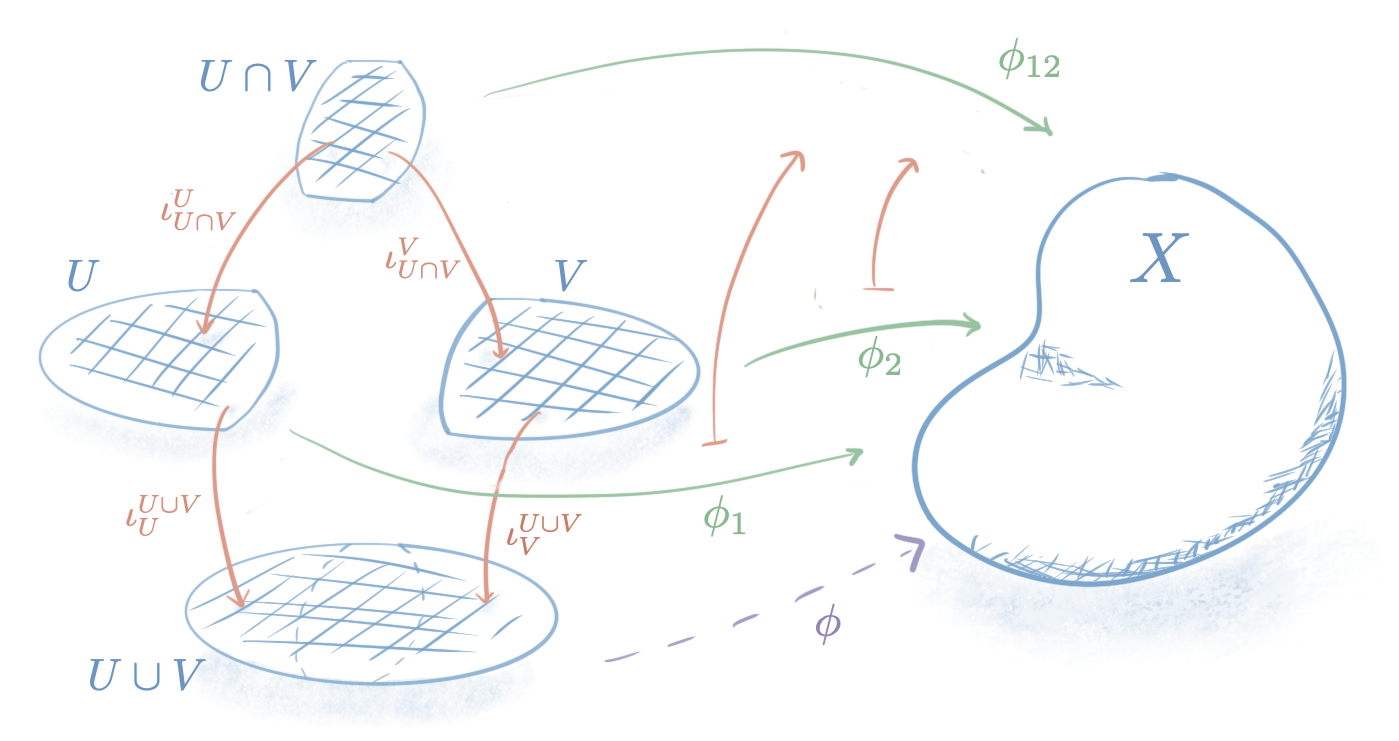}}
\vspace*{8pt}
\caption{A schematic illustration of the sheaf/glueing property. If two plots $\phi_1, \phi_2$ coincide on the intersection of their domains, there must be a plot $\phi$ defined on the union of the domains of $\phi_1$, $\phi_2$ whose restrictions coincide with them.}\label{fig:sheaf}
\end{figure}

The category \( \Eucl \) has additional structure. Namely, we can cover any open set $U$ using families $\{ U_i  \subset U\}_{i\in I}$ of open subsets, $U = \cup_{i\in I} U_i$. Then, if we can probe our smooth set with plots whose domains $U_i$ are the elements of some open cover, we should be able to assemble these plots into one whose domain is the union of the cover, provided they agree on the overlaps of the cover. More precisely, if \( \phi_1 \in X(U) \) and \( \phi_2 \in X(V) \) are two plots, if it happens that when we restrict them to the intersection \( U \cap V \) they agree, i.e.
\begin{equation}
	X(\iota_{U\cap V}^U)(\phi_1) = X(\iota_{U\cap V}^V)(\phi_2),
\end{equation}
then there must exist a unique plot \( \phi \in X(U\cup V) \) that restricts to \( \phi_1 \) and \( \phi_2 \) on \( U \) and \( V \) respectively, i.e.
\begin{equation}\label{eq:glueing}
	X(\iota^{U\cup V}_U)(\phi) = \phi_1 \qquad X(\iota^{U\cup V}_V)(\phi) = \phi_1 \, ,
\end{equation}
with $\iota_A^B \colon A \to B$, denoting the canonical embedding of the subset $A \subset B$. 
The glueing property of local plots is represented in Fig. \ref{fig:sheaf}.

A presheaf that satisfies the gluing condition (\ref{eq:glueing}) is called a \emph{sheaf}, and the categories on which such gluing condition can be formulated, like $\mathsf{Eucl}$, are called \emph{sites}. More precisely, a site is a category equipped with a so called \emph{Grothendieck topology}, which singles out, for every object, all the families of objects that cover it, see for example \cite{Deligne1973} for a standard reference. 

So far we have only exhibited smooth sets as objects, but we still need to define what kinds of maps we wish to consider between them. The way we do this is via plots. Indeed, a morphism \( f \colon X \to Y \) of smooth sets will induce a family of maps of plots \( f_U \colon X(U) \to Y(U) \) for every \( U \in \Eucl \). And this family ought to be compatible with changes of domains, i.e. for any \( \phi \colon U \to V \), we get a commutative square
\begin{equation}\label{eq:natural}
	\begin{tikzcd}
		{X(V)} && {Y(V)} \\
		\\
		{X(U)} && {Y(U)}
		\arrow["{f_U}", from=1-1, to=1-3]
		\arrow["{X(\phi)}"', from=1-1, to=3-1]
		\arrow["{Y(\phi)}", from=1-3, to=3-3]
		\arrow["{f_V}"', from=3-1, to=3-3]
	\end{tikzcd}
\end{equation}
The assignment $U \mapsto f_U$ satisying (\ref{eq:natural}) is called a natural transformation among the functors $X$ and $Y$, thus a ``map''  \( f \) of smooth sets is a \emph{natural transformation} $f \colon X \to Y$.

We have now a category in our hands, namely the category of sheaves on the site \( \Eucl \), otherwise known as the category of smooth sets that we will denote as $\mathsf{SmoothSet}$, whose objects are smooth sets, i.e., functors on the site $\mathsf{Eucl}$, and whose morphisms are maps among smooth sets, that is natural transformations among functors. Categories of sheaves on a site are examples of \emph{topoi} (also called Groethendieck topos), which are very convenient contexts in which to do mathematics from the categorical point of view. But it is not clear at this point that the category of manifolds is indeed fully faithfully embedded into the smooth sets. We show this next.

\subsection{Manifolds as smooth sets}
Given a manifold \( M \) it should be clear what the sheaf $X_M \colon \mathsf{Eucl}^{\mathrm{opp}} \to \mathsf{Set}$ it defines is. Given some domain \( U \in \Eucl \), the set of probes $X_M(U)$ with domain \( U \) into \( M \) should be \( C^\infty(U,M) \). This assignment \( U \mapsto C^\infty(U,M) \) is certainly functorial and contravariant, and one can show, in fact, that it defines a sheaf. For reasons that will become clear in just a moment, we write \( y(M) \coloneqq X_M =  C^\infty(-, M) \). Such a sheaf we will call representable.

We can then ask what the morphisms of smooth maps are between \( y(M) \) and \( y(N) \), with \( M \) and \( N \) two given manifolds. Here is where a fundamental result in category theory comes into play, the Yoneda lemma. This tells us that the natural transformations between two representable morphisms are in one to one correspondence with the maps between the objects that represent them. For our purposes, we have the following isomorphism\footnote{We adopt the commonplace notational convention of writing \( \mathsf{C}(a,b) \) for the set of morphisms $\alpha \colon a \to b$, between \( a \) and \( b \) in the category \( \mathsf{C} \). But, conceding to tradition, we write \( C^\infty(M,N) \) instead of \( \Mfd(M,N) \)}
  \begin{equation*}
    \SmoothSet(y(M), y(N)) \cong C^\infty(M,N) . 
  \end{equation*}
And hence \( y \colon \Mfd \to \SmoothSet \) is the functor that witnesses this embedding, and it is called the \emph{Yoneda embedding}.

We ought to say that, strictly speaking, \( y(M) \) is not a representable functor, for a representable functor is of the form \( \mathsf{C}(-,x) \) for \( x \) an object of \( \mathsf{C} \), but not every manifold is an object of \( \Eucl \). However, because of the sheaf condition, sheaves on \( \Eucl \) are actually the same as sheaves on \( \Mfd \), so that we could have just as well introduced smooth sets as sheaves on manifolds, which would make \( y(M) \) indeed a representable functor.

\subsection{Smooth sets as a topos}\label{sec:topos}
We have already hinted at how smooth sets form a topos, and topoi are well behaved categories. More precisely, they are complete with respect to the operations of taking limits and colimits, in the categorical sense. These limits are taken pointwise. As an example, given two smooth sets \( X \) and \( Y \), their product \( X \times Y \) is the smooth set whose plots are defined by
\begin{equation}
	(X \times Y)(U) \coloneqq X(U) \times Y(U)
\end{equation}
where the product on the right is the cartesian product of sets.

As a consequence, smooth sets can be very pathological. For example, the Cantor set can be constructed as a smooth set, since there exist smooth functions whose zero sets are the Cantor sets, and zero sets are examples of pullbacks, which exist in \( \SmoothSet \). 

Let us give another example of how to construct a smooth set. Given manifolds \( M \) and \( N \), \( C^\infty(M,N) \) is a smooth set whose plots are
\begin{equation}
	C^\infty(M,N)(U) \coloneqq C^\infty(U\times M,N).
\end{equation}
This makes a lot of sense, since, \( U \)-parametrised families of functions from \( M \) to \( N \) are the same things as functions from \( U \times M \) to \( N \). When the set of maps between two objects in a category is not just a set but can be identified with an object in the category, one calls that object an \emph{exponential}. 
%%%%%%%%%%%
%%%%%%%%%%%

\subsection{Examples of smooth sets}\label{sec:examples}
We outline in the following a number of basic examples showing how smooth sets appear in various field theories, in some cases even before it was realised that this was the structure that was lying underneath.   Of course, smooth manifolds and spaces of functions among two manifolds are examples of smooth sets as was discussed in the previous section, so it would not come as a surprise that the space of smooth sections of a bundle is a smooth set too.

\begin{example} The sections of a vector bundle.   The space of smooth sections $\Gamma (E)$ of a vector bundle $\pi \colon E \to M$ is a smooth set.   It can be argued that the space of sections of a bundle is a subset of the space $C^\infty (M, E)$ and because general properties of the category $\mathsf{SmoothSet}$ discussed in the coming section, it is a smooth set itself.  However it is also simple to describe the functor of plots for such space.   Given an open set $U \subset \mathbb{R}^n$, we denote by $\Gamma_U(E)$ the set of sections $\sigma_u$, $u\in U$, defined by the family of maps $F\colon U \times M \to E$ such that $\pi\circ F = \mathrm{pr}_2$, $\mathrm{pr}_2 \colon U \times M \to M$ denotes the canonical projection in the second factor, by means of $\sigma_u (x) = F(u,x)$, $u \in U$, $x\in M$.   It is an easy exercise to check that the sets of sections $\Gamma_U(E)$ are $U$-plots and they satisfy the glueing condition (\ref{eq:glueing}).  

Notice that other spaces of sections of vector bundles can be treated in the same way like, for instance, the space of metrics on a given manifold which is again a smooth set.   

Thus the spaces of sections of vector bundles, typical in many constructions of classical field theories, are smooth sets.   In Sect. \ref{sec:bicomplex} we will complete this picture by considering the canonical variational bicomplex associated to a given bundle.
\end{example}

\begin{example} Gauge field theories.   An extremely interesting application of the ideas presented in this paper is provided by the description of gauge fields in terms of parallel transport or, even better, holonomy maps.   This fruitful idea has been exploited in the loop quantum gravity program and has its roots on seminal papers trying to describe gauge theories in terms of the holonomies of connections (see, for instance, \cite{Wu1975,Gu1980,HongMo1986,Barrett1991}).     In fact J.W. Barret proved \cite{Barrett1991} that there is a one-to-one correspondence between group homomorphisms satisfying a technical condition, $H \colon \mathscr{P}(M,x) \to G$, where $\mathscr{P}(M,x)$ denotes the group of thin homotopy classes of loops on a connected manifold $M$ based on the point $x$, and $G$ denotes a compact Lie group; and the space of principal connections on a given $G$-principal bundle over $M$ (with a chosen point $p_0$  over $x$).       The technical condition that was spelled in the original paper by Barret, condition H3 \cite[p. 1183]{Barrett1991}, amounts to assert that the group homomorphism $H$ is indeed a smooth map from a diffeological space into $G$.    

Let us extend the discussion in the original paper by Barrett to the current setting.   Given a manifold $M$ we can consider its path groupoid $\mathscr{P}(M)$ over $M$ whose morphisms are thin homotopy equivalence classes of smooth paths $\gamma \colon [0,1] \to M$.   A thin homotopy $H \colon \gamma \to \gamma'$, $\gamma \sim_{\mathrm{thin}} \gamma'$,  is a smooth homotopy from $\gamma$ to $\gamma'$ such that $\int_{[0,1]\times [0,1]} H^* \alpha = 0$ for any 2-form $\alpha$ on $M$.   We will denote thin homotopy classes of paths just as $[\gamma]$ and the source and target maps $s,t \colon \mathscr{P}(M) \to M$ are defined as usual by $s([\gamma]) = \gamma (0)$, $t ([\gamma]) = \gamma (1)$.   The composition is defined by means of $[\gamma] \circ [\gamma'] = [\gamma \circ \gamma']$, provided that $s([\gamma]) = t([\gamma'])$ and the composition $\gamma \circ \gamma' (s) = \gamma' (2s)$, if $0\leq s \leq 1/2$, and  $\gamma \circ \gamma' (s) = \gamma (2s -1)$, if $1/2\leq s \leq 1$.  The inverse of the path $[\gamma]$ is given by the path $[\gamma^\dagger]$, defined as $\gamma^\dagger (s) = \gamma (1 - s)$.  Clearly $[\gamma^\dagger] \circ [\gamma] = [x]$ and $[\gamma] \circ [\gamma^\dagger] = [y]$, with $x = \gamma (0) = s([\gamma])$ and $y = \gamma (1) = t([\gamma])$, and $[x]$ is the thin homotopy class of the constant path $\gamma_x (s) = x$ for all $s$.     

The groupoid $\mathscr{P}(M)$ is a smooth set, in fact it is a diffeological space.   Certainly, the space of functions $C^\infty([0,1], M)$ is a smooth set as it was discussed in the previous section and the category of smooth sets, as it will be discussed in the coming section, is closed under quotients, then $\mathscr{P}(M) = C^\infty([0,1],M)/\sim_{\mathrm{thin}}$, is a smooth set.    In any case it is not too difficult to construct a suitable family of plots for  $\mathscr{P}(M)$ exhibiting explicitly its smooth set structure.   
If we denote by $\pi_{\mathrm{thin}}\colon C^\infty(I, M) \to \mathscr{P}(M)$ the canonical projection map, then we define the plots $\tilde{f} \colon U \subset \mathbb{R}^n \to \mathscr{P}(M)$ as $\tilde{f} = \pi_{\mathrm{thin}}\circ f$, for $f \colon U \to C^\infty(I, M)$ any plot in $C^\infty(I, M)$.

Finally if $\pi \colon P \to M$ is a right $G$-principal bundle, then an action of the groupoid $\mathscr{P}(M)$ on the principal bundle $P$ is a groupoid homomorphism $R \colon \mathscr{P}(M) \to \mathrm{Aut}(P)$, where $\mathrm{Aut}(P)$ is the groupoid over $M$ whose morphisms are maps $\phi_{yx} \colon P_x \to P_y$ such that $\phi_{yx} (pg) = \phi_{yx}(p)g$ for any $g\in G$, $x,y\in M$, $p \in P_x = \pi^{-1} (x)$, and such that for any given $x,y,p$ the maps $G \to P$, $g \mapsto \phi_{yx}(pg)$ is smooth.    The isotropy group of the groupoid $\mathscr{P}(M)$ at the point $x\in M$ will be denoted by $\mathscr{P}(M,x)$ and consists of thin homotopy classes of loops on $M$ based at $x$.   Thus an action of the groupoids of paths on the principal bundle $P$ defines a group homomorphism from the isotropy group $\mathscr{P}(M,x)$ into the isotropy group at $x$ of $\mathrm{Aut}(M)$ that can be identified with the structural group $G$ of the bundle $P$ by choosing a reference point $p_0 \in P_x$.   The groupoid homomorphism $H \colon \mathscr{P}(M,x) \to G$ defined in this way corresponds to the holonomy of the connection associated to it because of Barret's correspondence along the corresponding loop or, in other words:
$$
H ([\gamma]) = \mathrm{Hol}(A_H) =  \mathrm{P} \exp \int_\gamma A_H \, ,
$$
with $A_H$ the principal connection on $P$ defined by the homomorphism $H$.    Note that the technical condition introduced by Barret on the homomorphism $H$ in order to guarantee that there is a one-to-one correspondence between homomorphisms $H$ and principal connections is just the condition that the map $H$ is a smooth map from the smooth set $\mathscr{P}(M,x)$ to $G$ or, in more general terms that the action of the groupoid of paths on $P$ is smooth as smooth sets.

\end{example}

\begin{example} A nice example of the use of diffeological structures in the theory of fields is provided by the description of the constraint algebra of general relativity as the Lie algebroid structure associated to a diffeological groupoid \cite{Blohmann2013}. In this case, the diffeological space to be consider consists of a groupoid of morphisms determined by a family of spacetimes.   We will present here the basics of the construction in a slightly different form that is essentially equivalent to that in the original paper.   Let us consider a Riemannian manifold $(\Sigma, g)$ of dimension $d$, and  a globally hyperbolic spacetime $(\mathscr{M},\eta)$, that is $\eta$ is a Lorentzian metric of signature $(-+\cdots+)$ on the smooth manifold $\mathscr{M}$ of dimension $m = 1 + d$ (typically $m = 3,4$) such that $\Sigma$ is a Cauchy hypersurface of $\mathscr{M}$, that is there is an embedding $i\colon \Sigma \to \mathscr{M}$, and $i^*\eta = g$, and the submanifold $i(\Sigma) \subset \mathscr{M}$ cuts every inextensible temporal curve at exactly one point.    We will say that two embeddings $i\colon \Sigma \to \mathscr{M}$, and $i' \colon \Sigma \to \mathscr{M}'$ are equivalent, if there is an isometry $\varphi \colon (\mathscr{M}, \eta) \to (\mathscr{M}',\eta')$, such that $\varphi \circ i = i'$.   The space of equivalence classes $[i]$, of embeddings $i \colon \Sigma \to \mathscr{M}$, called $\Sigma$-universes, is a smooth set denoted by $\mathscr{U}(\Sigma)$.    We can define the sheaf that assigns to any open set $U\subset \mathbb{R}^n$, the equivalence classes of smooth maps $f \colon U \times \Sigma \to \mathscr{M}$, such that for every $u\in U$, the map $f(u,\cdot) : = i_u \colon \Sigma \to \mathscr{M}$ is an embedding of $\Sigma$ as a Cauchy surface of $\mathscr{M}$ and $f$ is equivalent to $f'$, if there is an isometry $\varphi \colon \mathscr{M} \to \mathscr{M}'$, such that $\varphi \circ f = f'$.   Then, $\mathscr{U}(\Sigma)$ is a smooth set.

The groupoid whose Lie algebroid describes the constraint algebra of gravity is the groupoid of equivalence classes of pairs of embeddings $i_1,i_2 \colon \Sigma \to \mathscr{M}$ on the same spacetime, that is we define $\mathscr{G}(\Sigma) = \{ [i_1,i_2] \mid [i_1], [i_2] \in \mathscr{U}(\Sigma)\}$, with $(i_1,i_2)$ equivalent to $(i_1',i_2')$ if there is an isometry $\varphi \colon \mathscr{M} \to \mathscr{M}'$, such that $\varphi\circ i_a = i_a'$, $a = 1,2$.  The groupoid $\mathscr{G}(\Sigma)$ will be called the groupoid of evolutions.    The composition law is given by, $([i,i']) \circ ([i'',i''']) = ([i,i''])$, provided that $[i'] = [i'']$.   Again it is easy to show that $\mathscr{G}(\Sigma)$ is a smooth set.   Thus, our groupoid is a groupoid in the category $\mathsf{SmootSet}$.    Both, in \cite{Blohmann2013} and \cite{Glowacki2019}, it is shown that the groupoid of evolutions is actually a diffeological space.   We can concentrate on the subcategory $\mathscr{U}(\Sigma)$, determined by all equivalence classes of embeddings $i \colon \Sigma \to \mathscr{M}$, where the underlying smooth manifold $\mathscr{M}$ is fixed.   
\end{example}

\begin{example}
	Let us briefly discuss how the description of spaces as sheaves on a site is also a convenient way to describe supergeometry. Supermanifolds are the settings in which theories that exhibit supersymmetry take place --and even without supersymmetry, supergeometry appears naturally when describing fermionic fields. Supermanifolds are, roughly speaking, manifolds that have anticommuting --also called fermionic-- in addition to commuting --bosonic-- local coordinates. Already, however, it is not clear how to implement this precisely, for the algebra of smooth functions of any ordinary manifold is always commutative. The way to circumvent this is to declare that the local models for supermanifolds are the formal duals of \emph{supercommutative} algebras. More specifically, there is a contravariant functor from the site of Euclidean spaces to the category of commutative algebras, by assigning to each Euclidean space its algebra of functions,
	\begin{equation}
		C^\infty(-) \colon \mathsf{Eucl}^{\mathrm{op}} \to \mathsf{Alg}
	\end{equation}
	which acts on morphisms by precomposition. This functor is in fact fully faithful, which means that Euclidean spaces can be identified with their algebras of functions. So then, the Euclidean \emph{super}space $\mathbb{R}^{m\mid n}$, with $m$ bosonic dimensions and $n$ fermionic dimensions, is \emph{defined} as the \emph{super}commutative \emph{super}algebra --where super is to be understood as $\mathbb{Z}_2$ graded--
	\begin{equation*}
		\mathbb{R}^{m \mid n} := C^\infty (\mathbb{R}^m) \otimes \wedge^\bullet (\mathbb{R}^n)^\ast.
	\end{equation*}
	This then defines a subcategory of the category of supercommutative superalgebras, whose opposite we define as the category $\mathsf{sEucl}$ of super Euclidean spaces. With some care, one can equip this category with a Grothendieck topology, in such a way that sheaves with respect to it become super smooth spaces. 
\end{example}

\begin{example}\label{ex:non concrete}
	All of the previous examples are examples of what are called \emph{concrete} smooth sets, otherwise known as \emph{diffeological spaces} (recall the definition of diffeological spaces provided in the introduction, Def. \ref{def:diffeological}). The points of a smooth set \( X \) are precisely the plots with domain a singleton, \( X(\ast) \). A diffeological space is one for which we have the following equality
	\begin{equation}
		X(U) \subseteq \Set(U, X(\ast))
	\end{equation}
	that is, the plots of \( X \) with domain \( U \) are indeed honest maps of sets from \( U \) seen as a set to the set of points of \( X \). It is not the case, however, that every smooth set is of this form. And one need not look very far to find examples. For a fixed \( p \in \mathbb{N} \), consider the following presheaf \( \Omega^p \colon \Eucl^\mathrm{op} \to \Set \) that sends each open domain \( U \) to the set of differential \( p \)-forms defined on it, and sends functions to their pullbacks. This is in fact a sheaf, and hence a smooth set. But it is not concrete. Indeed, the ``set of points'' of \( \Omega^p \) would be \( \Omega^p(\ast) \), but this contains only the trivial form if \( p>1 \). And yet, \( \Omega^p \) is far from being trivial. Non concrete smooth sets start to be relevant in the passage to super geometry, where the underlying set of points of two supermanifolds with the same even part is the same. 
\end{example}

%%%%%%%%%%%%%
%%%%%%%%%%%%%

\section{Geometry on the topos of smooth sets}\label{sec:geometry}
The category of smooth sets has very nice properties.   In fact the category of sheaves $\mathsf{Sh}(\mathsf{C})$ on a site $\mathsf{C}$ form what is called a Groethendieck topos.  In particular $\mathsf{SmoothSet} = \mathsf{Sh}(\mathsf{Eucl})$.   Topoi have all the basic ingredients needed for doing mathematics.  More precisely, on a topos all fundamental resources which are present in the category of sets are available, indeed the category of sets is a topos itself, and all limits, colimits and exponentials always exist on a topos \cite{Lurie2009}.   

\begin{figure}\label{fig:limits}
\centerline{\includegraphics[width=10cm]{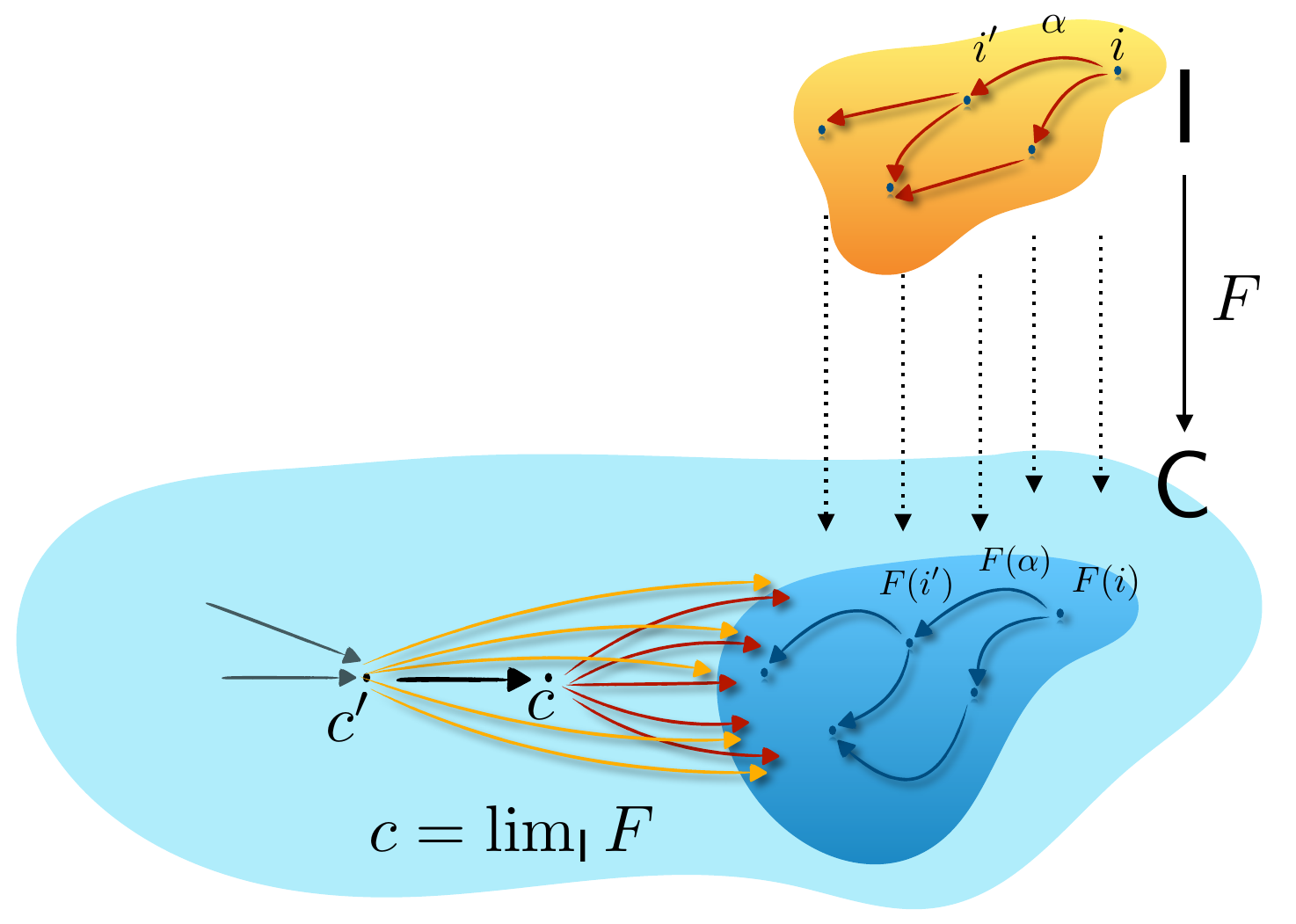}}
\vspace*{8pt}
\caption{A schematic illustration of the limit of a functor. The diagram $\mathsf{I}$ is represented (drawn) in the category $\mathsf{C}$ by the functor $F$ as $F(\mathsf{I})$, then the object $c =: \lim_\mathsf{I} F$ is the best covering of $F(\mathsf{I})$ in $\mathsf{C}$.}
\end{figure}

Limits and colimits of functors in categories are easily described in pictorial terms.   Let $F \colon \mathsf{I} \to \mathsf{C}$ be a functor from the category $\mathsf{I}$ into the category $\mathsf{C}$.   We may think of $F(\mathsf{I})$ as a realisation of the ``image'' $\mathsf{I}$ in the category $\mathsf{C}$.   We will also say that the functor $F$ is a diagram of type $\mathsf{I}$ in the category $\mathsf{C}$, then the limit of $F$, denoted $\lim_\mathsf{I} F$, is an object $c$ in the category $\mathsf{C}$ that ``captures'' the image $F(\mathsf{I})$ as best as possible, that is, on one side the set of morphisms $c \to F(i)$, cover consistently $F(\mathsf{I)}$ in the sense that $c \to F(i') = (F(i)\to F(i')) \circ (c\to F(i))$, for any morphisms $i \to i'$ in $\mathsf{I}$, and on the other,  $c$ is the ``best object covering'' $F(\mathsf{I})$ in the category $\mathsf{C}$.  The ``best object covering'' $c$ is formally expressed as the universal property that states that if $c'$ is another object covering the image $F(\mathsf{I})$ then there exists a morphism $c' \to c$ such that $(c\to F(i)) \circ (c' \to c) = c' \to F(i)$ for every object $i$ in the category $\mathsf{I}$\footnote{A compact way of expressing the previous ideas is by saying that $\lim_\mathsf{I} F$ is a final object in the subcategory of natural transformations from constant functors $\mathsf{I}_c \to F$, where $\mathsf{I}_c \colon \mathsf{I} \to \mathsf{C}$ is the constant functor that sends any $i$ to $c$.}.   See Fig. \ref{fig:limits} for a pictorial description of the limit of  a functor.  Colimits of functors are defined in a similar way by reversing the arrows.   

Thus, for instance, if we consider the category $\mathsf{I}$ defined by a diagram like 
\begin{equation}\label{eq:inverse}
i_0 \leftarrow i_1 \leftarrow i_2 \leftarrow \cdots \leftarrow i_k \leftarrow \cdots \, ,
\end{equation}
 (we can replace the previous diagram by any directed set $(\mathcal{I}, \prec )$), then the limit of a functor $F \colon \mathsf{I} \to \mathsf{C}$ becomes the so called inverse limit of the sequence:
\begin{equation}\label{eq:Finverse}
F(i_0) \leftarrow F(i_1) \leftarrow F(i_2) \leftarrow \cdots \leftarrow F(i_k) \leftarrow \cdots \, .
\end{equation}
Because an unfortunate conflict of terminologies ``limits'' in the categorical sense correspond to ``inverse limits'' in the set theoretical jargon and, conversely, ``colimits'' correspond to ``direct limits'' in set theory.   Thus the limit corresponding to diagram (\ref{eq:Finverse}) will be denoted $\varprojlim_k F$, or $\varprojlim_k F(i_k)$.

As it was mentioned before, we can launch a differential geometry program on topoi emulating the standard presentation of differential geometry in sets, that is reproducing the notion of smooth manifolds on the current topos by introducing a ring-like object $\mathcal{R}$ that will play the role of real numbers, and proceeding accordingly.   This program is known under the name of synthetic differential geometry and has been proposed as the natural way to geometrize the theory of quantum fields \cite{Schreiber2023}.      We will depart from this approach and will concentrate on the resources provided by smooth spaces to deal with geometric structures in the formulation of theories of fields.     In fact, the general strategy for doing that will consist of a two step procedure:  first, given a smooth space $X$, define our geometrical structure  consistently on plots and, second, use the sheaf property to glue them.   We will exhibit an explicit use of this strategy in the next section.

In this section however we will concentrate on one of the most fundamental geometrical structures in the category of manifolds: the tangent functor $T \colon \mathsf{Mfd} \to \mathsf{Mfd}$, that assigns to any manifold $M$ its tangent bundle $TM$, and to any smooth map $f \colon M \to N$, its tangent map $Tf \colon TM \to TN$.   The tangent functor plays a critical role in most geometrical constructions on manifolds and we would like to understand whether or not it is possible to extend it to the category of smooth sets. As it was discussed before the category of smooth manifolds is embedded naturally in the category of smooth sets by means of the Yoneda embedding $y \colon \mathsf{Mfd} \to \mathsf{SmoothSet}$, $M \mapsto y(M) = C^\infty(-,M)$. We can then ask the question of whether there exists a functor \( \hat{T} \colon \SmoothSet \to \SmoothSet \) that extends \( y \circ T \), i.e. such that \( \hat{T} \circ y = y \circ T \). There is a well known construction in category theory which provides us with exactly one such functor, namely the \emph{left Kan extension of \( y \circ T \) along \( y \)}\footnote{In general, the left Kan extension of a functor \( F \colon A \to B \) along a functor \( K \to A \to C \) is a functor \( \mathrm{Lan}_K F \colon C \to B \) that is universal in some specific sense, see \cite{Riehl2016} for details}. Because of the nice completeness properties of smooth sets, this extension always exists and fits into the diagram \ref{diag:Kan_ext}.
\begin{figure}
\begin{equation}
\begin{tikzcd}
	{\mathsf{Mfd}} && {\mathsf{Mfd}} \\
	\\
	{\mathsf{SmoothSet}} && {\mathsf{SmoothSet}}
	\arrow["T", from=1-1, to=1-3]
	\arrow["y"', from=1-1, to=3-1]
	\arrow["{y \circ T}", from=1-1, to=3-3]
	\arrow["y", from=1-3, to=3-3]
	\arrow["{\hat{T} = \mathrm{Lan}_y (y \circ T)}"', dashed, from=3-1, to=3-3]
\end{tikzcd}
\end{equation}
\caption{Diagram describing the construction of the tangent functor $\widehat{T}$ in the category $\mathsf{SmoothSet}$ as a left Kan extension.}\label{diag:Kan_ext}
\end{figure}

With this lift of the tangent functor to smooth spaces in hand, one can ask in what sense it is a reasonable tangent structure. In other words, is there a natural projection \( \hat{T}X \to X \), for \( X \) a smooth space? Do the fibers of this projection have a vector space structure? The answer in general is no, because these structures are defined in terms of natural transformations, and while they can be lifted using the machinery of Kan extensions, the properties they satisfy will generally not survive. This precise question is investigated by Blohmann in \cite{Blohmann2023a}. There, a subcategory of smooth sets, called \emph{smooth spaces}, is introduced such that \( \hat{T} \) restricts to a tangent structure on it. The condition of elasticity is quite technical, but roughly speaking it captures the property of being able to identify a vector space of tangent directions consistently from point to point, allowing for its dimension to jump. The subcategory of smooth sets that satisfy these conditions are called \emph{elastic spaces}. As it turns out, $C^\infty(M,N)$ is an elastic space and in that case one has,
$$
\hat{T}(C^\infty (M,N)) = C^\infty (M, TN)
$$
This fact is very commonly taken for granted in the usual discussion of the geometry of spaces of fields, where a tangent vector $\delta \phi$ to a field $\phi \colon N \to M$ is just a vector field on $N$ along the map $\phi$, that is a map $\delta\phi \colon M \to TN$, such that $\tau_N \circ \delta \phi = \mathrm{id}_N$, with $\tau_N \colon TN \to N$ the canonical projection as discussed in the introduction. But in the language of smooth sets this fact can be formalised and no arbitrariness needs to be introduced in the definition of the tangent space.   As it was pointed before, an alternative road to the construction of the tangent functor is provided by synthetic differential geometry, which is the approach proposed by Schreiber \cite{Schreiber2023} and developed to treat fermionic fields in Lagrangian fields theories in \cite{Giotopoulos2025b}.

A similar strategy can be used to extend Cartan calculus from the category of manifolds to the category of smooth sets, i.e., we denote the $\Omega_{\mathrm{dR}} \colon \mathsf{Mfd}^{\mathrm{op}} \to \mathsf{dgAlg}$, the Cartan sheaf that associates to any manifold $M$ its graded exterior algebra $\Omega^\bullet (M)$, which is an object in the category $\mathsf{dgAlg}$ of differential graded algebras. Then the left Kan extension along the Yoneda embedding allows us to define a functor $\widehat{\Omega}_{\mathrm{dR}} \colon \mathsf{SmoothSet} \to \mathsf{dgAlg}$ as $\widehat{\Omega}_{\mathrm{dR}}  = \mathrm{Lan}_y \Omega_{\mathrm{DR}}$ (see diagram \ref{diag:DR}).

\begin{equation}\label{diag:DR}
\begin{tikzcd}
	{\mathsf{Mfd}^\mathrm{op}} && {\mathsf{dgAlg}} \\
	\\
	{\mathsf{SmoothSet}^\mathrm{op}}
	\arrow["{\Omega_\mathrm{dR}}", from=1-1, to=1-3]
	\arrow["y"', from=1-1, to=3-1]
	\arrow["{\hat{\Omega}_\mathrm{dR} = \mathrm{Lan}_y \Omega_\mathrm{dR}}"', from=3-1, to=1-3]
\end{tikzcd}
\end{equation}

%%%%%%%%%%%%%
%%%%%%%%%%%%%

\section{Case study: the variational bicomplex}\label{sec:bicomplex}
\subsection{The infinite jet bundle as a smooth set}

Let us use the ideas introduced in the previous sections to see how they can aid us in dealing with the subtleties that come with working with infinite dimensional objects. In particular, we will outline the specific case of the so called infinite jet bundle $J^\infty\pi$, already presented in the introduction to this paper.

Jet bundles of fibre bundles are very natural objects in field theories, for they are the correct space on which to define differential operators on sections, i.e. maps that depend on the value of a section and its derivatives at some point. Most of the field theories of physical relevance are of finite order (second order more generally), so that one can get away with working on jet bundles of finite order which are finite dimensional smooth fiber bundles. Nevertheless, there is an advantage that comes from moving to the infinite jet bundle, not just because it supports theories of all orders, but more importantly because its tangent bundle comes with a natural splitting into vertical and horizontal distributions which makes its complex of differential forms into a bicomplex.  Most of the fundamental elements of a field theory (a Lagrangian, conserved currents, symmetries), can be identified as forms of a given bidegree in this bicomplex, known as the \emph{variational bicomplex}. Of course, this all comes at a price, namely the fact that the infinite jet bundle is not a finite dimensional manifold, so that it is not immediately clear how one should do differential geometry on it. It is, however, a smooth set, as we will explain below, so that we may use some of the techniques mentioned prior to guide is in making precise what is meant by vector fields and differential forms on it.

Let us briefly remind the reader of the definition of the infinite jet bundle. Given a bundle $\pi \colon E \to M$, one introduces the family of jet bundles $\pi^k \colon J^k \pi \to E$ as the space of equivalence classes up to order $k$ of germs of sections on $M$.  Given $(x^\mu, u^a)$ adapted coordinates on the bundle \( E \), $x^\mu$ being local coordinates on the base $M$ and $u^a$ local coordinates along its fibers, one can construct adapted coordinates on each of the \( J^k \pi \) bundles of the form $(x^\mu, u^a, u_\mu^a, u_{\mu\nu}^a, \ldots, u_{\mu_1 \cdots \mu_k}^a )$. Given some section \( \phi \colon U \subseteq M \to E \) around a point \( x \), the coordinates \( u^a_{\mu_I} \) are given by
  \begin{equation}
		u^a_I([\phi_x]) = \partial_I (u^a \circ \phi)(x)
  \end{equation}
for a multiindex \( I = (\mu_1, \ldots, \mu_l) \) of size less than or equal to \( k \). These are well defined precisely because points of \( J^k \pi \) are equivalence classes of sections of \( \pi \) which agree up to order \( k \). 

There are natural projections \( \pi^k_l \colon J^k\pi \to J^l\pi \), $k \geq l$, which assemble into a a so-called \emph{projective or inverse system} (recall the discussion at the beginning of Sect. \ref{sec:geometry} on limits and the specific example (\ref{eq:Finverse}) ):
\begin{equation}\label{eq:Jk}
E= J^0\pi \leftarrow J^1\pi \leftarrow J^2\pi \leftarrow \cdots \leftarrow J^{k-1}\pi \leftarrow J^k \pi \leftarrow \cdots .
\end{equation}
If \( J^\infty\pi \) would exist as a manifold, it would sit at the right most end of this sequence, and it would have projections to each of the finite order jet bundles. In categorical terms, it would be the limit of the diagram \ref{eq:Jk}, and its points would be sequences \( (\xi_0, \xi_1, \xi_2, \ldots,\xi^k, \ldots) \) such that \( \pi^l_k(\xi^l) = \xi^k \). That is, just like a point in the fiber \( J^k\pi \) over \( x \in M \) can be thought of as representing a Taylor polynomial centered at \( x \), a point in the infinite jet bundle of \( \pi \) would represent a formal Taylor series centered at \( x \). 

Of course, as we already hinted at before, the infinite jet bundle is not a manifold, since if it were it would have to have infinite dimension. Nevertheless, we can use the Yoneda embedding \( y \colon \Mfd \to \SmoothSet \) to transform the diagram in \ref{eq:Jk} into a diagram of smooth sets, and since the category of smooth sets is a topos, its limit will once again be a smooth set. It is this object that we call the infinite jet bundle \( J^\infty\pi \). In many references that introduce the infinite jet bundle, it is defined rather vaguely or treated purely formally (in \cite{Saunders1989} it is referred as a \emph{convenient fiction}), and of course it remains a purely formal object if we are restricted to the category of manifolds, but we need only broaden our context to smooth sets to see that it exists as an object of its own right and its introduction is rather straightforward. 

Let us explicitly state what plots into \( J^\infty \pi \) are. By the universal property that defines it, a smooth map into \( J^\infty \pi \) is the same thing as a family of maps into each of the \( J^k\pi \) that commute with the projections \( \pi^l_k \). That is, the datum of a plot into \( J^\infty\pi \) is exactly a sequence \( (\phi^0, \phi^1, \phi^2, \dots, \phi^k, \dots) \) with each \( \phi^k \) a plot of \( J^k\pi \) and \( \pi^k_l \circ \phi^k = \phi^l \). Equivalently, because the Yoneda embedding preserves limits, we may rephrase this fact as the following equality
\begin{equation}
	C^\infty(U, J^\infty \pi) = C^\infty(U, \varprojlim_k J^k\pi) = \varprojlim_k C^\infty(U, J^k\pi) 
\end{equation}
so that the set of plots with domain \( U \) into \( J^\infty\pi \) is precisely the limit of the projective system of plots into each \( J^k \).  

\subsection{Smooth function and tangent vectors on the infinite jet bundle}
As a first step towards constructing the variational bicomplex, we first compute the set of smooth functions \( J^\infty\pi \). By definition, we have
\begin{equation}
	C^\infty(J^\infty\pi) = \SmoothSet(J^\infty \pi, y(\R))
\end{equation}
i.e. the smooth functions on \( J^\infty\pi \) are just the maps of smooth sets from \( J^\infty\pi \) to \( \R \), seen as a smooth set via the Yoneda embedding. Now, we are trying to identify maps \emph{out of} \( J^\infty \pi \), so its defining universal property is not immediately helpful. It is shown in \cite{Giotopoulos2025, Khavkine2017} that in fact this set is precisely the set of \emph{locally finite order} functions. That is, a function (of plain sets) \( f \colon J^\infty\pi \to \R \) is smooth if and only if, there is a neighbourhood \( U \subseteq J^\infty\pi \) around each \( \xi \in J^\infty\pi \) such that \( f \) factors like
\begin{equation}
	\begin{tikzcd}
		{	J^\infty\pi \vert_U} && {   \R} \\
		\\
		{J^k\pi \vert_U}
		\arrow["{f \vert_U}", from=1-1, to=1-3]
		\arrow["{\pi^\infty_k \vert_U}"', from=1-1, to=3-1]
		\arrow["{f^k_U}"', from=3-1, to=1-3]
	\end{tikzcd}
\end{equation}
for some smooth (in the usual manifold sense) \( f^k_U \in C^\infty(J^k\pi, U) \). In other words, a smooth function out of the infinite jet bundle is one that, locally around each point, only depends on derivatives up to a finite order. This stands in contrast to the definitions given in \cite{Saunders1989, Anderson1992, Blohmann2023}, where smooth functions on the infinite jet bundle are defined as those that factor through a jet of finite order \emph{globally}. This definition is not correct if one is treating \( J^\infty \pi \) as a smooth set, in the sense that it does not capture the smooth maps from \( J^\infty\pi \) to \( \R \). Nevertheless, it can be defined in categorical terms as follows. We can apply the functor \( C^\infty(-) \colon \Mfd^{\mathrm{op}} \to \Alg \) to the diagram \ref{eq:Jk}. Because of the contravariance of the functor, the projections become injections and their direction is reversed, so that we obtain what is called a \emph{direct or injective system}, 
\begin{equation}
	C^\infty(J^0\pi) \rightarrow C^\infty(J^1\pi) \rightarrow C^\infty(J^2\pi) \rightarrow \cdots \rightarrow C^\infty(J^{k-1}\pi) \rightarrow C^\infty(J^k\pi) \rightarrow \cdots .
\end{equation}
The algebra of smooth functions of the infinite jet bundle would then sit at the right most end of the diagram, and it would have injections coming into it from each of the terms. In categorical terms, we would take the colimit of the diagram\footnote{This colimit exists because we are taking it in the category of algebras.} (which, as it was already mentioned in Sect. \ref{sec:geometry}, by some unfortunate choices of terminology is also sometimes called direct or injective limit). But, as one can compute, the colimit of this diagram is the algebra of functions which are \emph{globally} of finite rank,
\begin{equation}
	C^\infty_\mathrm{glob}(J^\infty\pi) \coloneqq \varinjlim_k C^\infty(J^k\pi).
\end{equation}
This is only a subalgebra of the smooth functions (in the sense of smooth sets) on \( J^\infty \pi \). 

A similar phenomenon occurs when attempting to define the tangent bundle of \( J^\infty \pi \). On the one hand, we can use the Kan extension of the tangent functor, as previously explained, and obtain the smooth set \( \hat{T}J^\infty\pi \). But we could also apply the tangent functor to \ref{eq:Jk}, followed by the Yoneda embedding \( y \colon \Mfd \to \SmoothSet \) to obtain a projective system of smooth sets:
\begin{equation}
y(TJ^0\pi) \leftarrow y(TJ^1\pi) \leftarrow y(TJ^2\pi) \leftarrow \cdots \leftarrow y(TJ^k\pi) \leftarrow \cdots 
\end{equation}
Once again, because smooth sets form a topos, this diagram has a limit, \( \varprojlim_k y(TJ^k\pi) \), which would sit at the right most edge, and come equipped with projections to each of the \( y(TJ^k\pi) \). Points of this space are sequences \( (X^0, X^1, \dots, X^k, \dots) \) with \( X^k \in TJ^k\pi \) and \( T\pi^k_l (X^k) = X^l \). If we write \( (x^\mu, u^a_I \) for the local adapted coordinates on each of the finite order jet bundles, we can write an element of \( \varprojlim_k y(TJ^k\pi) \) as a formal series
\begin{equation}
X = a^\mu\frac{\partial}{\partial x^\mu} + \sum_{ |I| \geq 0 } b_I^\mu \frac{\partial}{\partial u^a_I}.
\end{equation}

The immediate question is then what is the relation between these two objects. We know what the elements of \( \varprojlim_k y(TJ^k\pi) \) look like, but at this point it is no clear what the points of \( \hat{T}J^\infty \pi \) are. Nevertheless, we can show there exists a smooth map \(\hat{T}J^\infty\pi   \to \varprojlim_k y(TJ^k\pi) \). This follows from the universal property that defines \( \varprojlim_k y(TJ^k\pi) \). Indeed, we can apply \( \hat{T} \) to all of the projections \( \pi^\infty_k \colon J^\infty\pi \to y(J^k\pi) \) (we write explicitly the Yoneda embeddings) to obtain maps
\begin{equation}
	\hat{T}\pi^\infty_k \colon \hat{T} J^\infty\pi \to \hat{T} y(J^k\pi)
\end{equation}
but because \( J^k\pi \) is a manifold, it is true that \( \hat{T}y(J^k\pi) = y(TJ^k\pi) \). Hence we have a family of maps \( \hat{T}J^\infty \pi \to y(TJ^k\pi) \), which of course commute with the finite order projections, so, by the universal property, these must all factor through a unique map \( \hat{T}J^\infty\pi \to \varprojlim_k y(TJ^k\pi) \). In general, these two constructions of the tangent bundle will differ, and which is the correct choice will in large part depend on the specifics of the problem, nevertheless, structural relations can be derived by categorical arguments.

\subsection{Differential forms on the infinite jet bundle}
Finally we turn to the complex of differential forms on \( J^\infty \pi \). Just like previously, there are two possible definitions, one using the Kan extension outlined before, the other using the definition of \( J^\infty\pi \) as a limit, and they both lead to slightly different objects, depending on whether or not their rank is locally or globally finite. In any case, both constructions share the fundamental property of having two different gradings, which makes it into a bicomplex. There are various ways of introducing this bigrading, for example by showing that \( TJ^\infty\pi \) (for the reader's preferred choice of tangent bundle) has a canonical horizontal distribution, or that \( \Omega^\bullet(J^\infty\pi) \) has a distinguished ideal. But more natural when seeing \( J^\infty\pi \) as a smooth set is as follows. There is an evaluation map
\begin{equation}
	\mathrm{ev}_\infty \colon \Gamma(\pi) \times M \to J^\infty\pi
\end{equation}
that, given a section of \( \pi \) and a point in the base evaluates the section and all of its derivatives at that point. This is a map of smooth sets. One can show that when taking pullbacks this becomes an inclusion (see \cite{Giotopoulos2025} for more details):
\begin{equation}
	\mathrm{ev}_\infty^\ast \colon \Omega^\bullet(J^\infty\pi) \to \Omega^\bullet(\Gamma(\pi) \times M). 
\end{equation}
The complex of \( \Gamma(\pi) \times M \) is bigraded in an obvious way, and then this grading can be transported along \( \mathrm{ev}_\infty^\ast \) back to \( \Omega^\bullet(J^\infty\pi) \). This bicomplex is referred to as the \emph{variational bicomplex}. The degree which comes from the factor of \( M \) is called the \emph{horizontal degree}, whereas the one that comes from \( \Gamma(\pi) \) is called vertical. The forms in horizontal degree higher than the dimension of the base \( M \) must vanish, while there are forms with arbitrary vertical degree.
\begin{equation}
	\begin{tikzcd}
		\vdots & \vdots && \vdots \\
		{\Omega^{2,0}(J^\infty\pi)} & {\Omega^{2,1}(J^\infty\pi)} & \cdots & {\Omega^{2,\dim M}(J^\infty\pi)} \\
		{\Omega^{1,0}(J^\infty\pi)} & {\Omega^{1,1}(J^\infty\pi)} & \cdots & {\Omega^{1,\dim M}(J^\infty\pi)} \\
		{\Omega^{0,0}(J^\infty\pi)} & {\Omega^{0,1}(J^\infty\pi)} & \cdots & {\Omega^{0,\dim M}(J^\infty\pi)}
		\arrow[no head, from=2-1, to=1-1]
		\arrow[from=2-1, to=2-2]
		\arrow[no head, from=2-2, to=1-2]
		\arrow[from=2-2, to=2-3]
		\arrow[from=2-3, to=2-4]
		\arrow[no head, from=2-4, to=1-4]
		\arrow[from=3-1, to=2-1]
		\arrow[from=3-1, to=3-2]
		\arrow[from=3-2, to=2-2]
		\arrow[from=3-2, to=3-3]
		\arrow[from=3-3, to=3-4]
		\arrow[from=3-4, to=2-4]
		\arrow[from=4-1, to=3-1]
		\arrow[from=4-1, to=4-2]
		\arrow[from=4-2, to=3-2]
		\arrow[from=4-2, to=4-3]
		\arrow[from=4-3, to=4-4]
		\arrow[from=4-4, to=3-4]
	\end{tikzcd}
\end{equation}
The study of the variational bicomplex is extremely rich. Conservation laws, equations of motion and symmetries can all be stated in terms of forms in it, and the study of its cohomology solves the inverse problem for a set of differential equations, i.e., where they are actually the Euler-Lagrange equations of some Lagrangian theory, which was in fact one of the main initial motivation for its introduction (see, for instance, \cite{Deligne1999,Blohmann2023,Giotopoulos2025,Anderson1992} and references therein for detailed accounts). The exploration of the properties of the variational bicomplex from the perspective of smooth spaces will be continued in subsequent works.

% \section{Higher geometry of field theories}\label{sec:higher}
% \red{say something here if there paper is not too long!}

%%%%%%%%%%%%%
%%%%%%%%%%

\section*{Acknowledgments}
%%%%%
The authors acknowledge financial support from the Spanish Ministry of Economy and Competitiveness, through the Severo Ochoa Programme for Centres of Excellence in RD (SEV-2015/0554), the MINECO research project  PID2020-117477GB-I00, and Comunidad de Madrid project TEC-2024/COM-84 QUITEMAD-CM.
Arnau Mas is supported by a fellowship from ``La Caixa'' Foundation (ID 100010434), fellowship code LCF/BQ/DR23/12000033.
%%%%

\section*{ORCID}
\noindent Alberto Ibort - \url{https://orcid.org/0000-0002-0580-5858}

\noindent Arnau Mas - \url{https://orcid.org/0000-0003-0532-0938}

\bibliographystyle{ws-gm}
\bibliography{refs}

% \bibitem{Ga8*} K. Gawedzki.   Gauge fields in 2D. Commun. Math. Phys. (198*).
%
% \bibitem{Wi9*} E. Witten.  Equivariant integration 2D Yang-Mills. J. Geom. Phys. (199*).
%
% \bibitem{At**}  M. Atiyah, Index theorem and the space of loops (199*)
%
% \bibitem{Ib0*} A. Ibort.   Convexity formulas and index theorem (200*).
%
% \bibitem{Do8*} S. Donaldson.   Topology in 4D and gauge theories (198*).

% \end{thebibliography}

\end{document}